\documentclass[a4paper]{jpconf}
\usepackage{graphicx}
\usepackage{epstopdf}
\begin{document}
\title{Systematic Uncertainties at the T2K Experiment for Oscillation Measurements}

\author{A. Kaboth for the T2K Collaboration}

\address{Department of Physics, Imperial College London, London, SW7 2AZ, UK}

\ead{a.kaboth@imperial.ac.uk}

\begin{abstract}
The T2K experiment is a long-baseline neutrino oscillation experiment with the ability to  measure precisely $\nu_{\mu}\rightarrow\nu_e$ and $\nu_{\mu}\rightarrow\nu_{\mu}$ oscillations. Control of systematic uncertainties---dominated by flux and cross section uncertainties---is critical for the precision of these measurements. An analysis of charged current $\nu_{\mu}$ interactions at the T2K near detector significantly reduces these uncertainties, from 26.9\% to 3.0\% for the current $\nu_{\mu}\rightarrow\nu_e$ measurement at T2K.
\end{abstract}

\section{Neutrino Oscillation}

Neutrino oscillations are governed by the $3\times3$ Pontecorvo-Maki-Nakagawa-Sakata~\cite{MNS,Pont3} mixing matrix and parameterized by two mass-squared differences, $\Delta m^2_{21}$ and $\Delta m^2_{32}$;  three mixing angles, $\theta_{12}$, $\theta_{23}$, and $\theta_{13}$; and a complex CP-violating phase, $\delta_{CP}$. In recent years, the mass-squared differences and the mixing angles have all been measured to be non-zero~\cite{PhysRevD.86.010001}. However, the phase $\delta_{CP}$ remains unconstrained, and the precision of the measurements of the mixing angles must be improved in order to investigate this phenomenon. 

A critical factor in improving oscillation measurements is the control of systematic uncertainties, especially as more powerful neutrino sources are available and statistical uncertainty is thus reduced. Because much remains uncertain in the modeling of neutrino sources and cross sections, modern neutrino experiments use two detectors to constrain their systematic uncertainty by measuring the flavor content of the source close to the source, before neutrinos have had time to oscillate. This work will focus on the method and results of this technique for constraining systematic error in the context of the long-baseline Tokai-to-Kamioka (T2K) experiment. 

\section{The T2K Experiment}

The T2K experiment~\cite{Abe:2011ks} is a long-baseline experiment starting with a conventional muon neutrino beam of 30~GeV protons incident on a graphite target, produced at the J-PARC facility in Tokai, Japan.  The beam is characterized by a near detector complex, 280m downstream of the beam, consisting of an on-axis detector, INGRID~\cite{Abe2012},  and an off-axis detector, ND280. Oscillation measurements are made using the Super-Kamiokande (SK) detector~\cite{Fukuda:2002uc}, also off-axis from the beam, 295~km from the target. Using an off-axis beam allows for a narrowly focused beam at the maximal oscillation probability, and reduces uncertainty from higher energy interactions.

\begin{figure}[htbp] 
   \centering
   \includegraphics[width=0.4\textwidth]{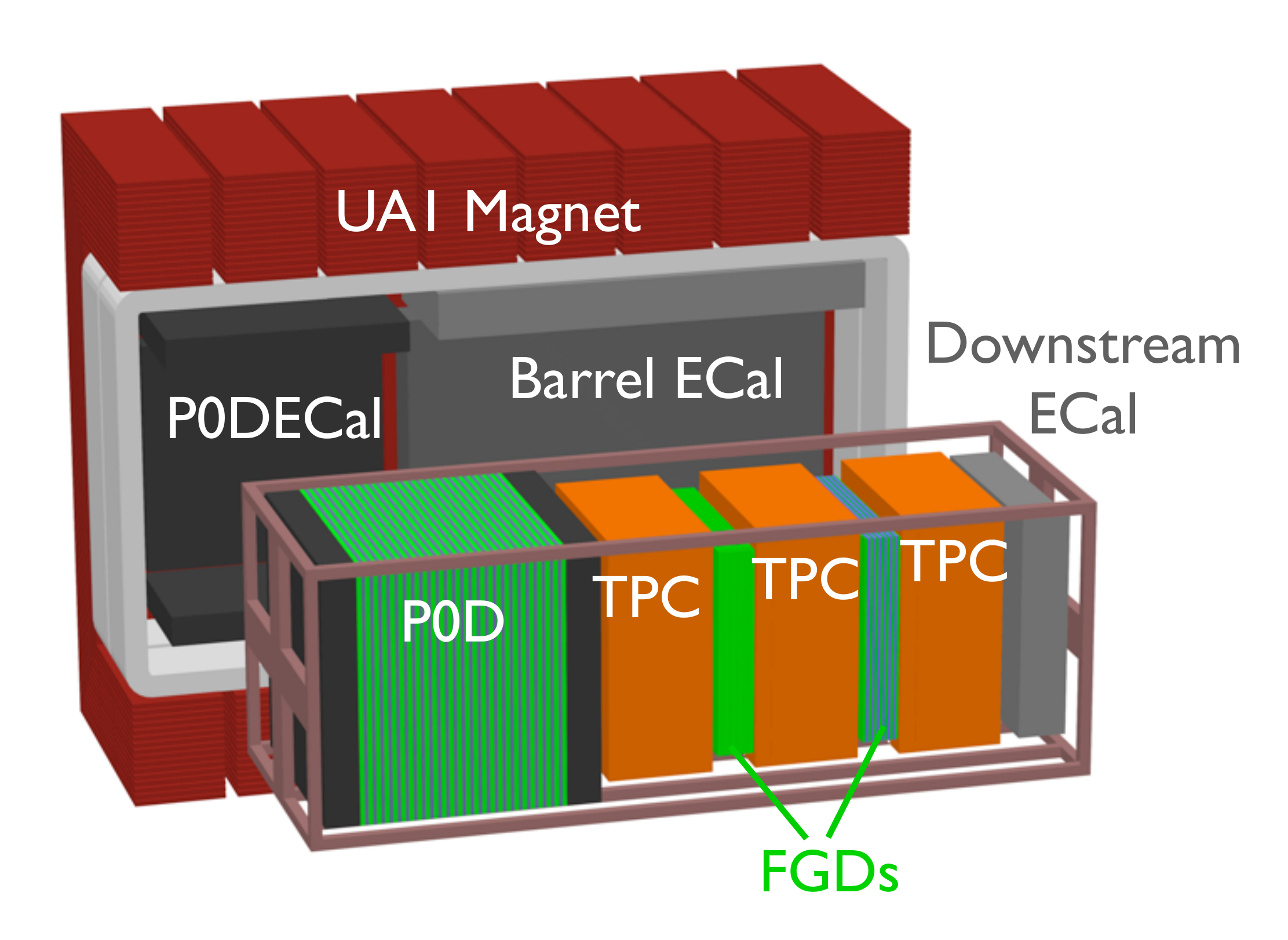} \hspace{2pc}%
\begin{minipage}[b]{14pc} \caption{ \label{fig:ND280}The T2K off-axis near detector, ND280. Shown in orange are the time projection chambers (TPCs), and in green, the fine grained detectors (FGDs). Also shown is the $\pi^0$ detector (P0D)~\cite{Assylbekov201248}, the electromagnetic calorimeters~\cite{allan2013electromagnetic}; and the UA1 magnet yoke and solenoid coils with muon range detector~\cite{smrd}, surrounding the inner detectors.}
\end{minipage}
\end{figure}

The ND280 detector, shown in Fig.~\ref{fig:ND280}, is the primary detector for the work discussed here, and is a multi-purpose detector composed of several sub-detectors. There is a central tracker region composed of two fine-grained detectors (FGDs)~\cite{Amaudruz:2012pe} surrounded by three three time projection chambers (TPCs)~\cite{Abgrall:2010hi}. The primary target for neutrino interactions in this analysis are the fine-grained detectors (FGDs), which comprise planes of plastic scintillator bars---providing a carbon interaction target---arranged in alternating directions and read out with wavelength shifting fibers attached to MPPCs. For this analysis, only the upstream FGD was used, with a total fiducial mass of 914~kg.  The primary method of determining particle identification and kinematics are the TPCs, which measures the momentum through the track curvature in a 0.2~T magnetic field and the particle identification through the dE/dx energy deposition in the gas. 


%

\section{Beam and Related Systematics}

A full account of the T2K beam and beam simulation can be found in~\cite{PhysRevD.87.012001}. The neutrino flux is predicted by modeling the interactions of the 30~GeV proton beam on the graphite target using FLUKA2008~\cite{Ferrari:2005zk}. The fluxes of the outgoing pions and kaons are tuned using data from the NA61/SHINE experiment~\cite{Abgrall:2011ae,Abgrall:2011ts} in momentum and angle bins. The decay and propagation of the hadrons through the magnetic horns, decay volume, and beam dump of the hadrons are modeled with GEANT3~\cite{GEANT3} and GCALOR~\cite{GCALOR}. 

\begin{figure}[htbp] 
   \centering
   \includegraphics[width=0.45\textwidth]{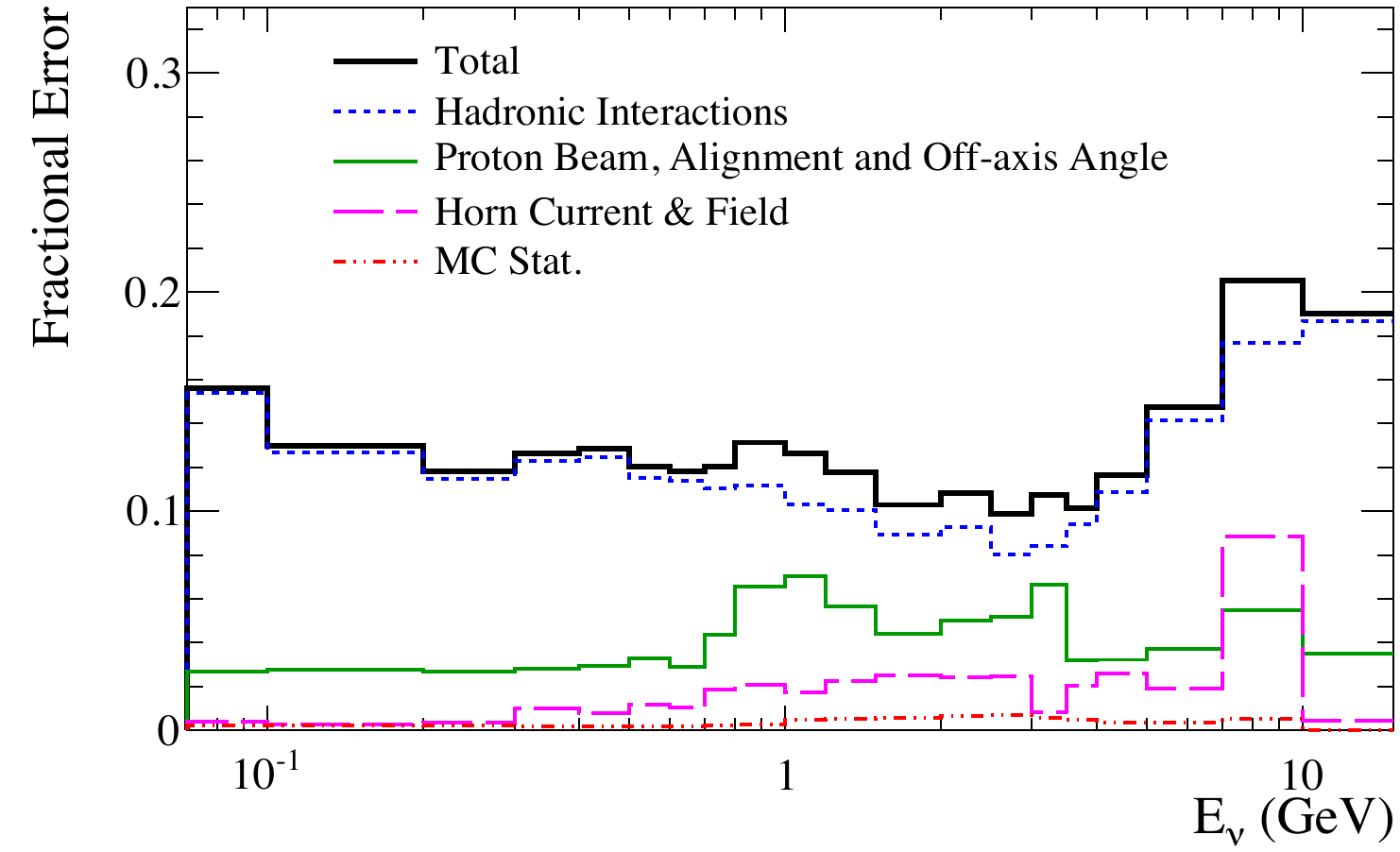}  \hspace{2pc}%
\begin{minipage}[b]{14pc}    \caption{  \label{fig:fluxuncertaintycontrib}Contributions to the flux uncertainty from various sources for the $\nu_{\mu}$ beam component at ND280. The dominant uncertainty comes from the tuning applied from NA61/SHINE data.}
\end{minipage}
\end{figure}

Systematic uncertainty is calculated on the beam from five sources: the proton beam, monitored in the proton beam line; the horn current uncertainty; alignment uncertainties on the target and horn; the beam direction, which is measured by the INGRID detector and beam line muon monitors; and the hadron production uncertainties from the NA61/SHINE tuning. These uncertainties are evaluated in bins of energy at ND280 and SK for the four neutrino species of the beam: $\nu_{\mu}$, $\bar{\nu}_{\mu}$, $\nu_e$, and $\bar{\nu}_e$.  Of these uncertainties, the largest contributor is the hadron production uncertainty, as shown in Fig.~\ref{fig:fluxuncertaintycontrib}. The uncertainties are provided to analyses as a covariance matrix, which allows the analysis at ND280 to constrain the flux uncertainties at SK. 

\section{Cross Section Model and Related Systematics}

Neutrino interactions at T2K are simulated using the NEUT neutrino interaction generator~\cite{Hayato:2009}. At the typical energies of the T2K beam, the dominant charged current (CC) interaction mode is charged current quasi-elastic (CCQE), with significant contribution from charged current single resonant pion production (CC1$\pi$). Because neutrino interactions at ND280 and SK are on different nuclei---carbon and oxygen, respectively---there are parameters that can be shared between the two detectors, relating to the underlying physics of neutrino interactions, as well as parameters that pertain only to one detector or the other, relating to the nuclear model. 

The most prominent of the first category of parameters are the two axial masses, $M_A^{QE}$ and $M_A^{RES}$, which govern the shape of the CCQE and CC1$\pi$ interactions. In addition to these, there are three CCQE normalization parameters in bins of energy of 0--1.5, 1.5--3.5, and $>3.5$~GeV, and two CC1$\pi$ normalizations for interactions of less than and greater than 2.5~GeV.  There is also a normalization of neutral current $\pi^0$ (NC$\pi^0$) production. This cross section is of particular importance, as NC$\pi^0$ events form a significant background for $\nu_{\mu}\rightarrow\nu_e$ oscillation analyses. These eight cross section parameters are the cross section parameters that are propagated in the covariance matrix between ND280 and SK.

Parameters that are evaluated separately at the two detectors include uncertainties on the binding energy and fermi momentum of nucleons from the fermi gas model of the nucleus. Additionally,  a systematic is evaluated on the difference between the fermi gas model and a spectral function model of the nucleus. Also included are uncertainties on the ratios of $\sigma_{\nu_{e}}/\sigma_{\nu_{\mu}}$ and $\sigma_{\bar{\nu}_{\mu}}/\sigma_{\nu_{\mu}}$ in oscillation analyses. 


The initial values and prior uncertainties for the cross section parameters are driven by data as much as feasible.  In particular, the value of $M_A^{QE}$ and the values of the single resonant pion production parameters ($M_A^{RES}$, CC1$\pi$ normalization, and NC1$\pi^0$ normalization) are determined by constraints from the MiniBooNE data~\cite{mb-ccqe,mb-cc1pi0}.


\section{ND280 Detector Systematics}

There are many sources of systematic uncertainty in the detector modeling at ND280: track kinematics determination---e.g., from uncertainties in the magnetic field; the efficiencies of each of the sub-detectors, as well as the matching efficiency between the sub-detectors; external background uncertainties---e.g. interactions from outside the FGD fiducial volume, event pileup, and interactions in the rock upstream of the detector; and monte carlo modeling---e.g., the modeling of the reinteraction rates of pions exiting primary neutrino interactions. These uncertainties are measured as far as is possible with external datasets (for example, cosmic muons used for efficiency and matching uncertainties) and the dominant uncertainties come from events outside the fiducial volume for muon momenta below 400~MeV/c and from pion reinteraction rates above 400~MeV/c. Typical detector uncertainties are smaller than flux and cross section uncertainties, with values of a few percent. 

\section{ND280 Event Selection and Parameter Constraints}

\begin{figure}[htbp] 
   \centering
   \includegraphics[width=\textwidth]{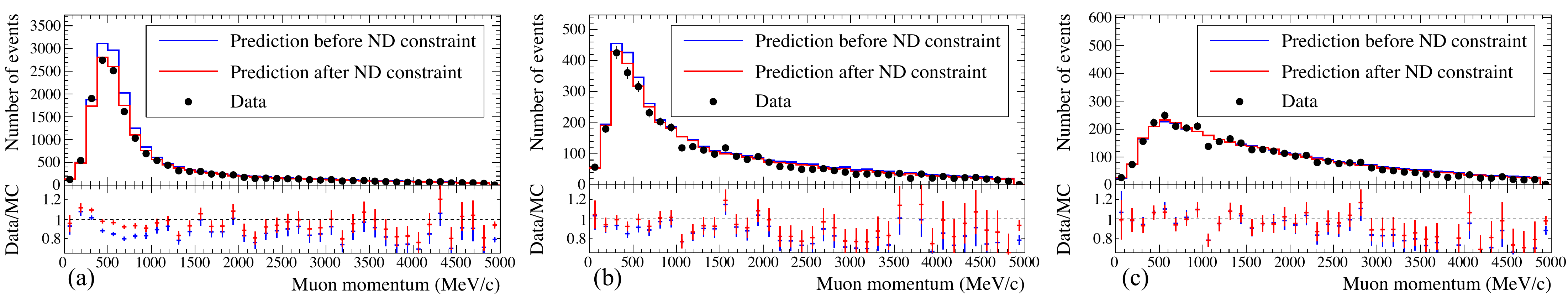} 
   \caption{The momentum distributions for the  (a) CC0$\pi$, (b) CC1$\pi^+$, and (c) CC other ND280 samples. Data is shown in black points, blue line shows the MC prediction before fitting the data, and red line shows the MC prediction after fitting the data.}
   \label{fig:CCmomdist}
\end{figure}

The event selection aims to constrain the flux and cross section systematics described above by selecting a CC $\nu_{\mu}$ interaction sample, and subdividing it according to the topology of the interaction. These sub-samples better constrain the cross section systematics on specific interaction modes. The selection begins by identifying muons as the highest momentum negative-curvature track emerging from the fiducial volume of the upstream FGD, with an energy deposit consistent with that of a muon in the next most downstream TPC. This sample is then divided into CC0$\pi$ (enhanced in CCQE interactions), CC1$\pi^{+}$ (enhanced in single resonant pion interactions), and CC other samples. The presence of a $\pi^{+}$ can be determined in three ways: a track emerging from the FGD to the TPC that has positive curvature and energy deposit consistent with that of a pion, a track fully contained in the FGD with energy deposit consistent with a pion, or a time-delayed energy deposit that is consistent with a $\pi^+\rightarrow\mu^+\rightarrow e^+$ decay chain.  Only the TPC method can conclusively determine the charge of a charged pion. Neutral pions are further identified with energy deposits in the TPC consistent with an electron. The CC0$\pi$ sample is required to have no identified pions of any charge in the event, the CC1$\pi^+$ sample must have one and only one identified positive pion, and the CC other sample has all remaining events. The purities are 72.6\%, 49.4\%, and 73.8\%, respectively.  These samples are fit, binned in momentum and angle with respect to the beam axis, using the flux, cross section, and detector systematic uncertainties described previously. Fig.~\ref{fig:CCmomdist} shows the momentum distributions before and after the fit for data and MC for all three subsamples. Table~\ref{tab:ND280rates} shows the number of events for data and pre- and post-constraint MC for the three subsamples. 

\begin{table}[htb]
\begin{center}
\small
\caption{Event rates for the ND280 detector, showing the number of data events, the number of MC predicted events before and after the fit to data.}
\label{tab:ND280rates}
\begin{tabular}{lccc}
                                     &  CC0$\pi$ & CC1$\pi$  & CC Other    \\  \hline\hline
Data                            &  16912     & 3936       & 4062       \\
Prediction (pre-fit)    &  20016.2  & 5059.4    & 4602.1        \\ 
Prediction (post-fit)  &  16802.5  & 3970.3    & 4006.0         \\ \hline\hline
\end{tabular}
\end{center}
\end{table}

\begin{figure}[htbp] 
   \centering
   \includegraphics[width=0.95\textwidth]{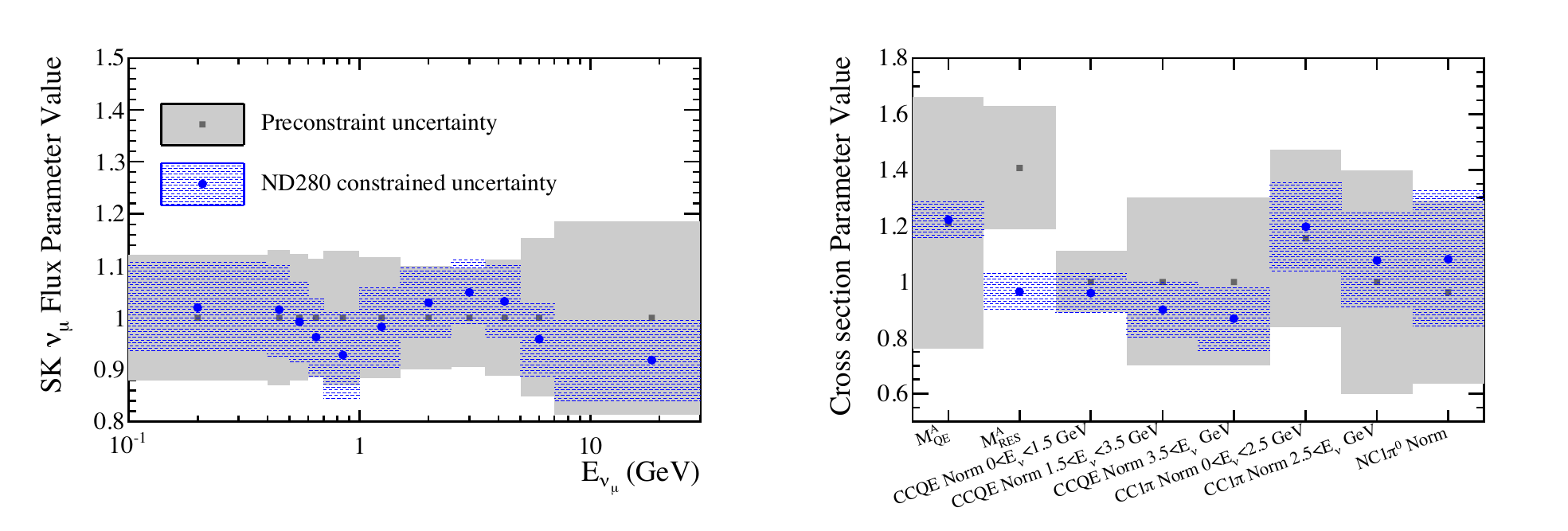} 
   \caption{The effect of the ND280 analysis on the parameter uncertainties propagated to oscillation analyses. Left: the impact on the flux normalizations for the muon neutrino flux at SK. Right: the impact on the cross section parameters. The value of the axial masses is in GeV; all other parameters are normalizations.}
   \label{fig:SKfluxconstraint}
\end{figure}

Fig.~\ref{fig:SKfluxconstraint} shows the impact of the constraint on the flux and cross section parameters propagated to SK. There is both significant reduction in the uncertainty on each parameter, and also significant shifts in the parameters. Most shifts are within the systematic error assigned to them before fitting the data, but the value of $M_{A}^{RES}$ moves significantly away from its prior value. This shift is primarily driven by the CC1$\pi^+$ sample; this sample is dominated by single resonant pion events, and, as shown in Table~\ref{tab:ND280rates}, the initial predicted value of the number of events is higher than the number of data events. 

The constraint from the ND280 data significantly reduces the error on the number of predicted events at SK for oscillation analyses, especially because there are negative correlations between the constraints on the propagated flux and cross section parameters. Table~\ref{tab:NSK} shows the magnitude of this effect, compared with other sources of error in the analysis; using the ND280 data reduces the impact of the propagated parameters from the largest source of error by far to the smallest source of error. 

\begin{table}[htb]
\begin{center}
\small
\caption{Percentage error on the number of events expected in a $\nu_{\mu}\rightarrow\nu_e$ oscillation analysis at SK, for a value of $\sin^2 2\theta_{13}=0.1$.}
\label{tab:NSK}
\begin{tabular}{lcc}
 Uncertainty Source & Without ND280 Constraint & With ND280 Constraint \\ \hline
Flux and Cross section from ND280 & 26.9\% & 3.0\% \\
Other Cross section & 7.5\% & 7.5\% \\
SK Detector & 3.5\% & 3.5\% \\
\hline
Total & 28.2\% & 8.8\% \\
\hline 
\end{tabular}
\end{center}
\end{table}

\section{Conclusions and Future Thoughts}

The sources of systematic uncertainty at the T2K experiment are many, and, before data constraints, can be quite large. The T2K near detector, ND280, has been effectively employed to select a sample of charged current muon neutrino interactions that tightly constrain these uncertainties. The division of the sample into subsamples that address different neutrino interaction modes is particularly useful. Further work to reducing these uncertainties will come from studying other interaction modes, $\nu_e$ interactions, and interactions on materials other than carbon at ND280.

\section*{References}
\bibliography{Syst}

\end{document}